\documentclass[final,5p,times,twocolumn]{elsarticle}

\usepackage{multirow} % 导入multirow宏包
\usepackage{amssymb}
\usepackage{amsmath}
\usepackage{enumitem}
\usepackage{natbib}
\journal{Neural Networks}

\begin{document}

\begin{frontmatter}



\title{Noisy Disentanglement with Tri-stage Training for Noise-Robust Speech Recognition}


\author[aff1]{Shuangyuan Chen} %% Author name
\author[aff1]{Shuang Wei}
\author[aff2]{Dongxing Xu}
\author[aff1,aff2]{Yanhua Long\corref{cor1}} 
\cortext[cor1]{Yanhua Long is the corresponding author (yanhua@shnu.edu.cn). 
This work is supported by the National Natural Science Foundation of China (Grant No.62071302).}

\affiliation[aff1]{organization={Shanghai Engineering Research Center of Intelligent Education and Bigdata, Shanghai Normal University},%Department and Organization
            city={Shanghai},
            country={China}}
            
\affiliation[aff2]{organization={Unisound AI Technology Co., Ltd.},%Department and Organization
            city={Beijing},
 %           postcode={}, 
 %           state={},
            country={China}}

%% Abstract
\begin{abstract}

To enhance the performance of end-to-end (E2E) speech recognition systems in noisy or low 
signal-to-noise ratio (SNR) conditions, this paper introduces NoisyD-CT, a novel tri-stage training framework 
built on the Conformer-Transducer architecture. The core of NoisyD-CT is a especially designed 
compact noisy disentanglement (NoisyD) 
module (adding only 1.71M parameters), integrated between the Conformer blocks and Transducer Decoder to 
perform deep noise suppression and improve ASR robustness in challenging acoustic noise environments.
To fully exploit the noise suppression capability of the NoisyD-CT, we further propose a clean representation consistency 
loss to align high-level representations derived from noisy speech with those obtained from corresponding clean speech. 
Together with a noisy reconstruction loss, this consistency alignment enables the NoisyD module 
to effectively suppress noise while preserving essential acoustic and linguistic features consistent 
across both clean and noisy conditions, thereby producing cleaner internal representations that enhance ASR performance. 
Moreover, our tri-stage training strategy is designed to fully leverage the functionalities of both the noisy disentanglement 
and speech recognition modules throughout the model training process, ultimately maximizing performance gains under noisy 
conditions. Our experiments are performed on the LibriSpeech and CHiME-4 datasets, extensive results demonstrate that 
our proposed NoisyD-CT significantly outperforms the competitive Conformer-Transducer baseline, achieving up to 25.7\% and 
10.6\% relative word error rate reductions on simulated and real-world noisy test sets, respectively, while maintaining or 
even improving performance on clean speech test sets. The source code, model checkpoint and data simulation 
scripts will be available at https://github.com/litchimo/NoisyD-CT.

\end{abstract}



%% Keywords

\end{frontmatter}



\section{Introduction}      
\label{sec1}
With the rapid advancement of technologies such as neural network architectures, automatic speech recognition (ASR) has achieved remarkable progress in recent years~\cite{chorowski2015attention}~\cite{li2021better}~\cite{kim2017joint}. Traditional ASR systems are primarily designed in a modular fashion, comprising components like feature extraction, acoustic modeling, and language modeling, relying heavily on manually crafted features and statistical modeling techniques~\cite{dong2018speechtransformer}~\cite{hinton2012deep}~\cite{martin2009speech}. However, the independent optimization of these modules poses limitations, making it challenging for the overall system to achieve optimal performance. With the rise of deep learning, end-to-end speech recognition methods based on neural networks have become mainstream, significantly improving performance~\cite{xiong2018microsoft}~\cite{li2022recent}. Nevertheless, in complex noisy environments or under low signal-to-noise ratio (SNR) conditions, end-to-end ASR models are highly susceptible to environmental noise, leading to signal distortion and substantial performance degradation~\cite{seltzer2013investigation}.

To address this issue, many studies have incorporated speech enhancement (SE) models as the front-end of automatic speech recognition (ASR) systems~\cite{weninger2015speech}. For the SE module, both traditional methods~\cite{scalart1996speech} and neural network-based approaches can be employed, with implementations possible in both the time domain~\cite{pandey2019new}~\cite{defossez2020real} and frequency domain~\cite{michelsanti2017conditional}~\cite{wang2020complex}. Additional, the SE module and the ASR module can be trained either independently~\cite{fujimoto2019one} or jointly~\cite{wang2016joint}. However, the training objectives of SE modules typically focus on improving speech quality, such as minimizing mean squared error (MSE) or enhancing signal-to-noise ratio (SNR). In contrast, ASR models are generally optimized based on word error rate (WER)~\cite{fan2020gated}. This mismatch between the objectives of the SE and ASR modules can lead to distortions in the speech processed by the SE module, thereby negatively affecting the performance of the ASR system. Consequently, while the SE-based approach has limitations, alternative methods that eliminate the SE module entirely and employ end-to-end models for improved noise robustness remain relatively unexplored.

To overcome these challenges and achieve superior performance with enhanced robustness, we propose the NoisyD-CT framework. Based on the Conformer-Transducer 
 (Conformer-T) architecture, which has been widely adopted as a state-of-the-art model in speech recognition, the NoisyD-CT framework has made three key innovations as follow:

\begin{enumerate}[label=\arabic*),left=1pt, labelsep=0.3em]
    \item Introduce a noisy disentanglement module to improve noise-robust ASR performance. Due to the limited performance of the Conformer-Transducer model in speech recognition under complex noisy conditions, we incorporate an additional noisy disentanglement module within the framework. Specifically, we introduce three lightweight modules (1.71M parameters) between the encoder and decoder of the Conformer-T. These modules perform deep noise suppression on the representations obtained from the encoder, aiming to generate as clean a representation as possible, which is then fed into the decoder. This enhancement significantly improves the overall noise robustness of the model.
    
    \item Propose a tri-stage training strategy. The tri-stage training strategy is designed to maximize the performance of each module in the whole NoisyD-CT, addressing the limitations of traditional single-stage training. Initially, each module is trained independently to ensure it effectively learns its specific function. Once the individual modules have been optimized, they are integrated and undergo joint training, refining their interactions to enhance overall ASR performance.
    
    \item Propose a clean representation consistency loss together with a noisy reconstruction loss to further enhance the NoisyD-CT. The original Conformer-T model is trained using only Connectionist Temporal Classification (CTC) loss and RNN-T loss. In our approach, we extend this by incorporating two additional losses, leveraging both clean representations and reconstructed noisy representations to guide the model training. This integrated strategy enhances ASR performance in noisy conditions while maintaining performance on clean speech tasks.
\end{enumerate}

The rest of this paper is organized as follows. Section \ref{sec2} presents the review of related works.
In Section \ref{sec3}, we briefly describe the fundamental of the Conformer-Transducer based ASR architecture. 
In Section \ref{sec4}, we introduce the proposed NoisyD-CT framework, including the model structure and tri-stage 
training strategy. The experimental setup and results are presented in Section \ref{sec5} and Section \ref{sec6}. 
Finally, we conclude our work in Section \ref{sec7}.

%% Use \subsection commands to start a subsection.
%\subsection{Introduction}
%\label{subsec1}

\section{Review of Related Works}
\label{sec2}

\begin{figure*}[!htbp]
    \centering
    \includegraphics[width=0.75\linewidth]{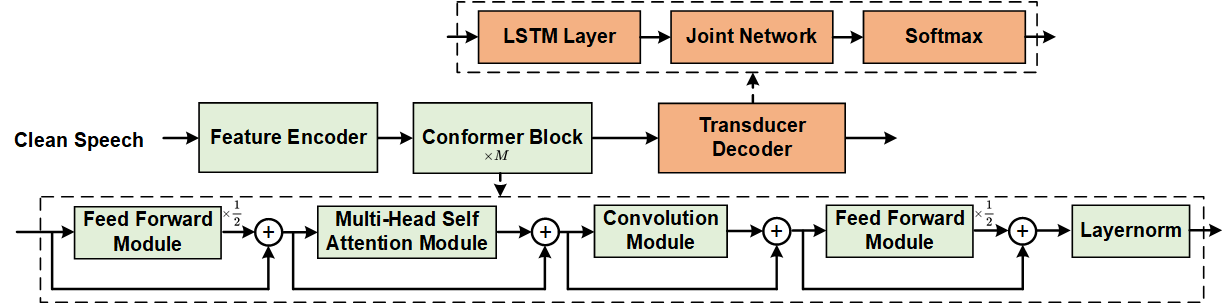}
    \caption{The framework of Conformer-Transducer.}
    \label{fig1}
\end{figure*}

Noise-robust speech recognition aims to enhance the performance of speech recognition systems in complex noisy environments or under low signal-to-noise ratio (SNR) conditions. This technology is particularly significant for real-world scenarios where environmental noise is prevalent~\cite{donahue2018exploring}. The challenges of noise-robust ASR include the diversity and complexity of noise types, interference and distortion in audio signals~\cite{pandey2021dual}, as well as the robustness and generalization capability of models. To address these challenges, previous research has primarily focused on two key areas: (i) SE-based noise-robust ASR; (ii) Self-supervised learning in noisy ASR.

\textbf{SE-based noise-robust ASR}: To improve the noise robustness of ASR systems and achieve better results under noisy conditions, a widely adopted approach is to incorporate a speech enhancement (SE) module~\cite{kinoshita2020improving}. This approach is particularly effective in challenging noisy conditions, such as scenarios with signal-to-noise ratios (SNR) approaching 0 dB, where speech intelligibility becomes critical for accurate recognition. To address this, some studies have proposed a cascaded framework, where the SE module operates as a front-end to enhances speech intelligibility before feeding the output into the ASR system~\cite{fujimoto2019one}. However, since the objectives of SE and ASR modules are not aligned and they are trained independently in some studies, issues such as speech distortion introduced by the SE module can negatively impact the performance of the downstream ASR system. Evidence from the DNS 2023 Challenge~\cite{dubey2024icassp} underscores this issue: while the TEA-PSE3.0 model achieved a high OVRL MOS score of 2.71, its word accuracy (WAcc) in speech recognition was only 0.761. This highlights a critical limitation that higher MOS, PESQ, and other speech enhancement subjective scores do not necessarily translate to higher speech recognition accuracy~\cite{10096838}.To bridge this gap, later research has explored joint training of SE and ASR models~\cite{liu2019adversarialenhancement}. By adopting strategies like multi-task learning~\cite{ma2021multitask}, these approaches maximize the correlation between the SE and ASR modules, allowing them to complement each other~\cite{li2023audiovisual}. Ultimately, these advancements enable SE modules to enhance speech intelligibility and contribute to improved ASR performance. However, challenges remain, as optimizing SE for intelligibility does not always align with ASR objectives, potentially introducing distortions that degrade recognition accuracy.

\textbf{Self-supervised learning in noisy ASR}: To enhance ASR performance under noisy conditions, another widely adopted approach is self-supervised learning (SSL). By leveraging large amounts of unlabeled noisy speech data, self-supervised learning can help build robust ASR systems. For example, to enhance noise robustness, Wav2vec-Switch~\cite{wang2022wav2vec} extends wav2vec2.0 (w2v2)~\cite{baevski2020wav2vec} by incorporating a contrastive loss that encodes raw noise into the contextual speech representations. Building on the same w2v2 baseline, ~\cite{zhu2022noise} proposes an enhanced w2v2 by minimizing the discrepancy between noisy and clean features, thereby obtaining more robust speech representations. Similarly, ~\cite{huang2022spiral} utilize a teacher-student framework to encode denoised representations extracted from noisy data. ~\cite{wang2022improving} introduces an additional auxiliary reconstruction module to further improve the noise robustness of learned SSL representations. Additionally, disentangled HuBERT (deHuBERT)~\cite{ng2023dehubert} incorporates a novel pair of auxiliary loss functions that promote noise invariance in HuBERT’s contextual representations. While self-supervised learning significantly improves ASR robustness in both known and unseen noisy conditions, it comes with challenges such as large-model size and high computational resource requirements. 
Moreover, without careful fine-tuning, the learned representations may not always align optimally with ASR objectives, potentially limiting their performance upper-bound.

Other studies have tried to integrate SE modules with self-supervised learning to enhance ASR robustness in noisy environments.\ For example,~\cite{ravanelli2020multi} proposed the PASE+, which incorporates an online speech distortion  module to contaminate the input signals with a variety of random disturbances. By applying various random perturbations to corrupt the input signal before self-supervised learning, PASE+ demonstrates strong performance in noisy environments.~\cite{zhu2023jointspeech} employs a dual-attention fusion module to combine SE’s ability to enhance speech quality with self-supervised pretraining’s noise robustness, mitigating information loss and speech distortion. However, this method is limited to offline ASR tasks and is unsuitable for real-time applications. To further refine integration, ~\cite{woo2024knowledge} introduce a knowledge distillation-based joint training method, which constructs a pipeline where the ASR and SE models are treated as teacher and student models, respectively. By converting their outputs into a unified acoustic token space, this method bridges their objective gap, improving linguistic information extraction and speech quality. Additionally,~\cite{hu2018ganaugmentation} combine SE modules with Generative Adversarial Networks (GANs)~\cite{goodfellow2014generative} to enhance ASR performance in both known and unknown noise environments. While these methods significantly improve ASR robustness in noisy conditions, they come with notable trade-offs: 1) The introduction of additional modules, such as dual-attention fusion or GAN-based frameworks, substantially increases the model’s parameter count, leading to higher computational complexity; 2) Although these approaches enhance performance in noisy environments, they often result in a significant performance drop when processing clean speech. This highlights the challenge of balancing robustness in noisy conditions with maintaining high accuracy in clean speech scenarios.

In this study, we focus on enhancing ASR performance in complex noisy environments while minimizing the increase in parameter count and ensuring no degradation on clean datasets. To achieve this, we propose NoisyD-CT, a novel end-to-end framework that leverages synthetically generated noisy data to improve ASR robustness under noisy conditions. By integrating noisy disentanglement methods and optimized training strategies, NoisyD-CT strengthens noise robustness without compromising system efficiency or performance on clean speech.

\section{Conformer-Transducer based E2E ASR}
\label{sec3}

\begin{figure*}[h]
    \centering
    \includegraphics[width=0.75\linewidth]{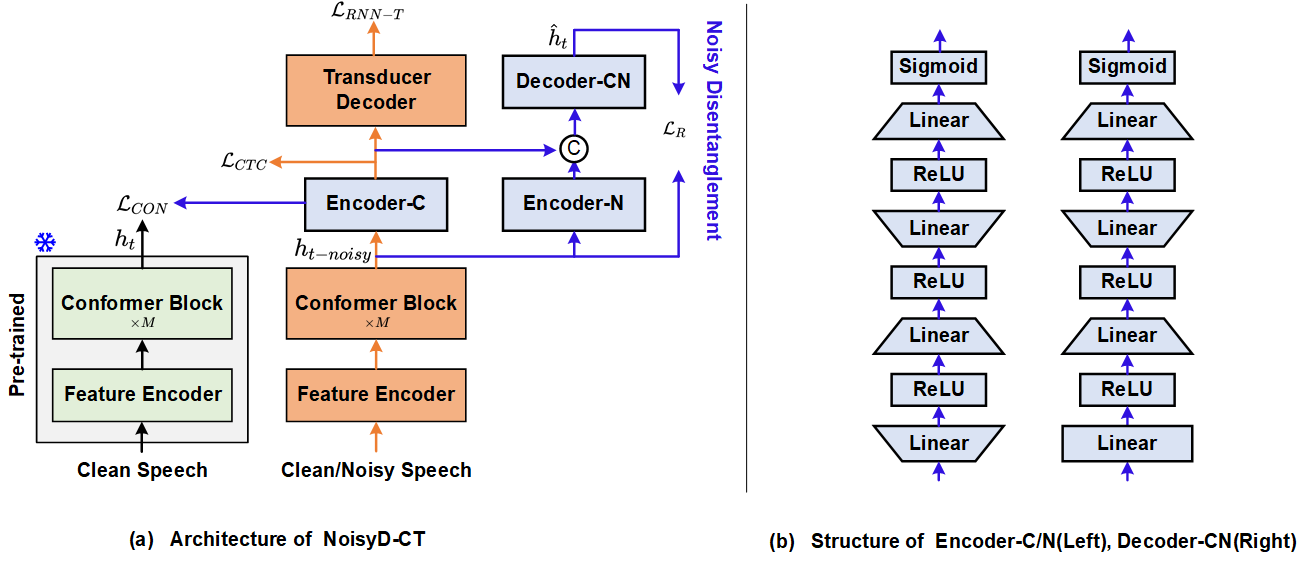}
    \caption{The framework of proposed NoisyD-CT E2E ASR. }
    \label{fig2}
\end{figure*}

In this paper, all our contributions are based on the Conformer-Transducer (Conformer-T), an end-to-end ASR model originally proposed in~\cite{gulati2020conformer}~\cite{zhang2020transformer}. The Conformer-T has demonstrated state-of-the-art performance across various ASR tasks by seamlessly integrating the Conformer's local feature modeling with the Transducer’s streaming processing capabilities. These advantages have made Conformer-T increasingly popular in recent end-to-end ASR systems, making it the ideal choice as the baseline model for this study. The whole architecture of Conformer-T is shown in Figure \ref{fig1}. As a standard transducer architecture, Conformer-T consists of a Feature Encoder, followed by $M$ Conformer blocks as the encoder, and a Transducer Decoder. 

\textbf{Feature Encoder}: The Feature Encoder consists of a stack of two-dimensional convolutional layers. We denote \(x=[x_1,x_2,...,x_T]\) as an acoustic feature input with $T$ frames, which are extracted from the raw audio clip and \(y=[y_1,y_2,...,y_U]\) as the corresponding transcription label sequence of length $U$. Feature Encoder downsamples acoustic feature $x$ into latent acoustic representations \(z=[z_1,z_2,...,z_t](t<T)\). The corresponding equation is as follows:
\begin{equation}
z = FeatureEncoder(x)
\tag{1}
\end{equation}

The Feature Encoder not only reduces the computational demand of the entire model but also highlights the essential information needed for automatic speech recognition.

\textbf{Conformer Blocks}: After the Feature Encoder, $M$ Conformer blocks, serving as the main part of the encoder, extract high-dimensional acoustic feature representations $h_t$ from latent acoustic representations $z$. Each Conformer block consists of the following components in sequence: a Feed Forward network(FFN) module, a Multi-Head Self Attention (MHSA) module, a Convolution (CONV) module, and a FFN module at the end. Furthermore, Layernorm and residual connections are applied to stabilize training and allow for deeper stacking of layers. The overall process is illustrated as follows:
\begin{equation}
\begin{aligned}
z_{FFN_1} &= z + \frac{1}{2} FFN(z) \\
z_{MHSA} &= z_{FFN_1} + MHSA(z_{FFN_1}) \\
z_{CONV} &= z_{MHSA} + CONV(z_{MHSA}) \\
z_{FFN_2} &= z_{CONV} + \frac{1}{2} FFN(z_{CONV}) \\
h_t &= Layernorm(z_{FFN_2})
\end{aligned}
\tag{2}
\end{equation}

The advantage of this architecture lies in its ability to learn long-range global interaction information from the attention module, while also effectively capturing local features through the convolutional module. Additionally, the two half-step feed-forward layers demonstrate superior performance in capturing complex patterns and dependencies within the input sequence.

\textbf{Transducer Decoder}: The LSTM layer, the joint network, and the Softmax layer form the Transducer Decoder together. After the Conformer blocks, the LSTM layer generates a high-dimensional label representation $h_u(u<U)$ using the previous non-blank label $y_{u-1}$ from the transcription label sequence $y$ as input. The joint network then combines the high-dimensional acoustic representations $h_t$ with the label representations $h_u$, and passes them through a Softmax layer to predict the probability of the next label.\ Moreover, RNN-T loss~\cite{graves2012sequence} is also employed. Given $\mathcal{A}(y)$, the set of all possible alignment $l$ with special blank token between input $x$ and output $y$,the loss function can be computed as the following negative log posterior: 
\begin{equation}
\mathcal{L}_{RNN-T} = -log \textit{P}(y|x) = -log \sum_{l\in \mathcal{A}(y)} \textit{P}(l|x)
\tag{3}
\end{equation}

Moreover, as in~\cite{jeon2021multitask}, we also train Conformer-T with an auxiliary CTC loss to learn frame-level representation. For a given acoustic input $x$ and a label sequence $y$, CTC finds all possible alignments $\mathcal{B}(x,y)$ by computing all potential paths, with the following calculation:
\begin{equation}
\mathcal{L}_{CTC} = -log \textit{P}(y|x) = -log \sum_{\pi \in \mathcal{B}(x,y)} \textit{P}(\pi|x)
\tag{4}
\end{equation}

So the overall loss of Conformer-T is defined as:
\begin{equation}
\mathcal{L}_{Conformer-T} = \mu \mathcal{L}_{CTC}+\gamma \mathcal{L}_{RNN-T}
\tag{5}
\end{equation}
where $\mu,\gamma \in[0,1]$ are tunable loss weights.

\section{\textbf{Proposed Methods}}
\label{sec4}

In this section, we provide a detailed description of the proposed NoisyD-CT ASR framework, which is designed to enhance the noise robustness of end-to-end speech recognition systems. The overall model architecture is presented in Section \ref{subsec1}. The noisy disentanglement module is detailed in Section \ref{subsec2}, and Section \ref{subsec3} describes the tri-stage training strategy including pre-training, noisy disentanglement training, and fine-tuning. 

\subsection{Architecture}
\label{subsec1}

To achieve robust recognition of noisy speech, we propose the NoisyD-CT E2E framework, as 
illustrated in Figure \ref{fig2}(a). This framework builds upon the traditional Conformer-Transducer 
architecture outlined in Section \ref{sec3}, with key components including a 
separate \textit{pre-trained Feature Encoder} and \textit{pre-trained Conformer blocks}, \textit{Conformer-Transducer} and 
the \textit{Noisy Disentanglement}. The structure of the pre-trained Feature Encoder and Conformer blocks 
follows the same design as their counterparts in the complete Conformer-T 
architecture. After being trained on a clean speech dataset, their parameters are fixed and used to 
generate high-dimensional clean representation for each corresponding noisy speech representation target. 
This alignment of high-level representations between noisy and clean speech provides a strong 
noise suppression guidance for the proposed noisy disentanglement module (NoisyD). 
Leveraging this alignment, the NoisyD module in Figure \ref{fig2}(a) performs a comprehensive 
disentanglement operation, which separates the high-dimensional noisy representations into distinct 
noise and clean speech components, effectively minimizing interference from noise 
and preserving the semantic information within the clean speech representation.
The NoisyD enhanced acoustic representations are then fed into a Transducer Decoder to produce the 
transcriptions. Furthermore, to ensure each module of the proposed NoisyD-CT retains its specific functionality 
while resolving misaligned objectives, we propose a tri-stage training strategy. 
This strategy enables the whole framework to jointly enhance the noise-robust ASR performance. 
It should be noted that during ASR inference, the \textit{pre-trained Feature Encoder}, \textit{pre-trained Conformer blocks}
and the \textbf{Encoder-N} and \textbf{Decoder-CN} of the NoisyD module in Figure \ref{fig2}(a) are all discarded. 

\subsection{Noisy Disentanglement}
\label{subsec2}

In real-world noisy environments, noise often disrupts the effective information carried by speech, leading to a significant decline in ASR performance, especially under low signal-to-noise ratio (SNR) conditions. To address this issue, the noisy disentanglement (NoisyD) is introduced into the overall NoisyD-CT framework. It aims to obtain clean representations from noise corrupted speech at the high-dimensional feature representation level, which is more robust to noise interference compared to processing at the raw speech feature level.

As illustrated in Figure \ref{fig2}(a), the noisy disentanglement module comprises three key components: 
\textbf{Encoder-C}, \textbf{Encoder-N}, and \textbf{Decoder-CN}. Additionally, the framework incorporates the clean representation consistency loss $\mathcal{L}_{CON}$ and the noisy speech reconstruction loss 
$\mathcal{L}_{R}$. These loss functions play a crucial role in ensuring a comprehensive disentanglement process, 
facilitating the extraction of  clean speech representations from noisy inputs.

Specifically, as shown in Figure \ref{fig2}(a), given $x$ as the clean speech's acoustic feature and 
$x_{noisy}$ as the acoustic features of clean/noisy speech. By replacing $x$ with $x_{noisy}$ in Eq.(1) and Eq.(2), 
we obtain the high-dimensional acoustic noisy representation $h_{t-noisy}$, which serves as the input to the NoisyD 
module. As demonstrated in Figure \ref{fig2}(b), the structures of \textbf{Encoder-C}, \textbf{Encoder-N}, and 
\textbf{Decoder-CN} each consist of four linear layers with activation functions. ReLU is used as the primary 
activation function, while a Sigmoid activation function is applied after the final linear layer. 
This structural design allows the modules to progressively learn complex features and extract the desired information 
while ensuring minimal additional model parameter overhead. The \textbf{Encoder-C} and \textbf{Encoder-N} share an 
identical architectural design but are tailored for distinct high-level representation extractions. Specifically,  
\textbf{Encoder-C} is responsible for extracting the purified clean representation \(\tilde{h}_{\text{clean}}\), which 
is primarily used as input to the Transducer Decoder for further decoding. On the other hand, \textbf{Encoder-N} is 
designed to extract the pure noise representation \(\tilde{h}_{\text{noisy}}\), which is then concatenated with the 
clean representation and fed into \textbf{Decoder-CN}. While \textbf{Decoder-CN} follows a similar structural design, 
it differs in the dimensionality of its first linear layer. Its primary function is to reconstruct the 
high-dimensional acoustic representation \(\hat{h}_{\text{t}}\), thereby indirectly 
facilitating the extraction of clean representations. The overall process unfolds as follows: 

\begin{equation}
\begin{aligned}
\tilde{h}_{\text{clean}} &= \text{Encoder-C} (h_{t-noisy}) \\
\tilde{h}_{\text{noisy}} &= \text{Encoder-N} (h_{t-noisy}) \\
\hat{h}_{\text{t}} &= \text{Decoder-CN} (Concat(\tilde{h}_{\text{clean}},\tilde{h}_{\text{noisy}}))
\end{aligned}
\tag{6}
\end{equation}

To achieve the noise suppression function of the NoisyD module, two MSE losses are introduced: 
the clean representation consistency loss $\mathcal{L}_{CON}$  and the noisy reconstruction loss $\mathcal{L}_{\text{R}}$. 
$\mathcal{L}_{CON}$ plays a crucial role by enforcing the alignment of high-level representations derived from 
noisy speech with those obtained from their target clean speech. This alignment enables the NoisyD module to 
suppress noise while preserving the essential acoustic and linguistic information that are consistent 
across both clean and noisy conditions. As a result, the whole model is guided to separate noise more effectively, 
leading to “cleaner” internal representations that enhance the downstream Transducer Decoder’s performance.
By minimizing the representation distance between $h_t$ and \(\tilde{h}_{clean}\), the $\mathcal{L}_{CON}$
is defined as follows:

\begin{equation}
\mathcal{L}_{CON} = \frac{1}{N} \sum_{i=1}^N \left(h^i_{\text{t}} - \tilde{h}_{clean}^i\right)
\tag{7}
\end{equation}
where $N$ is the batch size, $h_t$ represents the high-dimensional acoustic representation obtained from the 
clean speech input processed by the frozen \textit{pre-trained Feature Encoder} and \textit{Conformer blocks}, 
as detailed in Section \ref{sec3}.

The noisy reconstruction loss $\mathcal{L}_{\text{R}}$ promotes the extraction of pure noise 
representation \(\tilde{h}_{\text{noisy}}\) by \textbf{Encoder-N}, which does not contain valid linguistic and 
semantic content, by minimizing the distance between the reconstructed acoustic representation 
\(\hat{h}_{\text{t}}\) and the extracted high-dimensional acoustic representations $h_{t-noisy}$. 
This, in turn, indirectly enhances the noise suppression ability 
of \textbf{Encoder-C}. Formally, the $\mathcal{L}_{\text{R}}$ is defined as: 
\begin{equation}
\mathcal{L}_{R} = \frac{1}{N} \sum_{i=1}^N \left(\hat{h}^i_{t} - h^i_{t-noisy}\right)
\tag{8}
\end{equation}

With Eq.(5), the whole loss function of our NoisyD-CT E2E framework is then formulated as: 
\begin{equation}
\mathcal{L}_{NoisyD-CT} =\mathcal{L}_{Conformer-T} + \alpha \mathcal{L}_{CON} + \beta \mathcal{L}_R
\tag{9}
\end{equation}
where $\alpha,\beta \in[0,1]$ are the weighting parameters. 
$\mathcal{L}_{Conformer-T}$ is detailed in Section \ref{sec3}.

\subsection{Tri-stage Training Strategy}
\label{subsec3}

\begin{figure*}[h]
    \centering
    \includegraphics[width=0.75\linewidth]{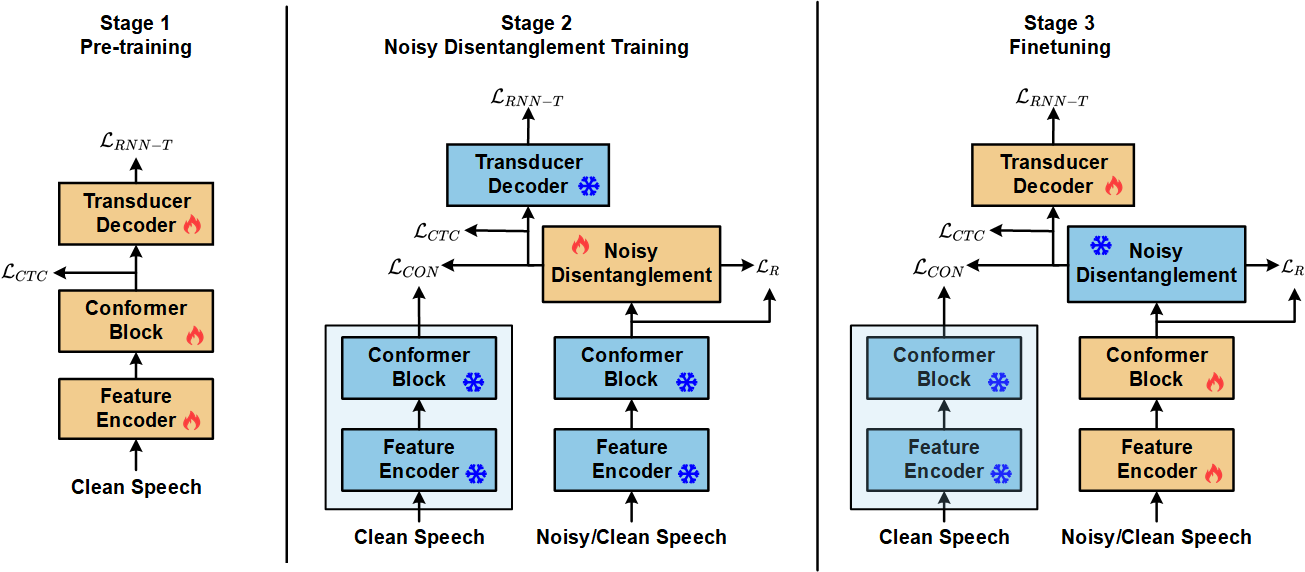}
    \caption{Block diagram of tri-stage training strategy on the NoisyD-CT framework.}
    \label{fig3}
\end{figure*}

In the proposed NoisyD-CT architecture, a single-stage training approach can introduce interference between the noisy 
disentanglement process and the core ASR task, often leading to degraded performance. To eliminate this issue and 
fully leverage each module’s functionality, we propose a tri-stage training strategy in this section. 
As shown in Figure \ref{fig3}, this strategy first isolates the Conformer-Transducer’s training to ensure robust speech recognition 
capabilities under clean conditions, then incrementally incorporates the noisy disentanglement module 
to refine noise suppression, and finally fine-tunes the entire network. 
By separating these learning objectives into distinct phases, the model more effectively disentangles 
noise while preserving linguistic information, ultimately maximizing the performance gains in noisy ASR scenarios.
The whole strategy is described in the following sections.

\textbf{Pre-training}: In this initial stage, a standard Conformer-Transducer model is trained on the clean speech 
training set with the standard objective of achieving state-of-the-art (SOTA) performance on clean speech. 
Specifically, only $\mathcal{L}_{CTC}$ and $\mathcal{L}_{RNN-T}$ are used for optimization, 
providing a robust set of initialization parameters for the subsequent stages. 
This approach not only reduces overall training time but also endows the Feature Encoder and Conformer blocks with the capability to extract 
high-quality acoustic representations $\hat{h}_{t}$. In addition, such a strong foundation can also guarantee the 
effectiveness of the proposed $\mathcal{L}_{CON}$ for noisy disentanglement and then ultimately enhancing ASR 
performance under noisy conditions.

\textbf{Noisy Disentanglement Training}: As shown in Figure \ref{fig3}, the second stage isolates the training of 
the noisy disentanglement (NoisyD) module from the rest of the network. All parameters for the Feature Encoder, 
Conformer blocks, and Transducer Decoder (previously obtained in the pre-training stage) are kept fixed, 
ensuring that the knowledge acquired from clean speech remains intact. As clearly shown in the figure, 
the clean speech and its simulated noisy speech are separately fed into the parallel structured 
pre-trained Feature Encoder and Conformer blocks, providing paired clean-noisy high-level representations for 
the implementation of $\mathcal{L}_{CON}$ in NoisyD training. The whole NoisyD module alone is 
then optimized using the overall model loss function of $\mathcal{L}_{NoisyD-CT}$, focusing on 
its noise suppression capabilities without adversely affecting the other components.

By preventing any noise-related gradient updates from propagating to the pre-trained modules, 
this stage preserves the acoustic and linguistic representations already learned. At the same time, 
the dedicated training of NoisyD highlights its ability to disentangle and suppress noise, 
producing an effective initialization for the next fine-tuning stage. As a result, the entire system 
can later be jointly refined with minimal interference, ultimately achieving superior recognition 
performance under noisy conditions.

\textbf{Fine-tuning}: In the final stage, only the backbone modules (the Feature Encoder, Conformer blocks, and Transducer 
Decoder) of standard Conformer-Transducer within the proposed NoisyD-CT framework are fine-tuned, 
while the parameters of all other blocks remain frozen, as obtained from the initial pre-training 
and noisy disentanglement stages. This fine-tuning stage is performed using all the complete set of paired 
clean-clean and clean-noisy training data, and it is optimized via the overall loss function $\mathcal{L}_{NoisyD-CT}$.
This fine-tuning stage is very crucial, because it adapts the main Conformer-Transducer network to 
effectively handle both clean and noisy inputs, thereby maximizing the benefits of the noise suppression achieved 
in Stage 2 while simultaneously boosting ASR performance.

While Stage 1 only uses clean data, it is important to note that both the noisy disentanglement training (Stage 2)
and the fine-tuning stage (Stage 3) utilize the entire paired dataset, comprising both 
clean-clean (the same speech recording) and clean-noisy training pairs. In Stage 2, when clean-clean pairs 
are input, the value of $\mathcal{L}_{CON}$ is theoretically near zero. Incorporating these pairs 
during model training can prevent the noisy disentanglement module from excessively suppressing 
the clean speech input, thus ensuring the good ASR performance on speech under both clean and noisy 
application scenarios. This tri-stage training design choice guarantees that, at inference, 
only the right part of Feature Encoder, Conformer blocks at the Conformer-Transducer backbone, 
\textbf{Encoder-C} from the NoisyD module, and the Transducer Decoder are employed, 
achieving an good balance between noise suppression, ASR performance, and model complexity.

\section{Experimental Setup}
\label{sec5}

In this section, we provide a detailed description of the datasets and experimental configuration used 
during the system training, validation and evaluation.

\subsection{Datasets}
\label{subsec5}

We use the open-source LibriSpeech dataset~\cite{panayotov2015librispeech} as our clean speech corpus, 
utilizing the train-clean-100 and train-clean-360 subsets for training, the dev-clean and dev-other sets for 
validation, and the test-clean and test-other sets for evaluation. For noisy speech simulation, noise samples are 
sourced from the CHiME-4 Challenge dataset~\cite{vincent2017analysis}, which comprises real-world noise recordings captured with six-channel distant microphone arrays and close microphones in four distinct environments: buses(BUS), 
cafes(CAF), pedestrian areas(PED), street intersections(STR). To generate synthetic noisy speech, we randomly select one-channel noise samples from the 
CHiME-4 dataset and mix them with LibriSpeech clean recordings at randomly chosen signal-to-noise ratios (SNRs) ranging 
from –5 dB to 15 dB. This process produces synthetic noisy speech data that are paired one-to-one with the clean 
speech, forming noisy-clean training pairs (train-noisy-100 vs. train-clean-100, train-noisy-360 vs. train-clean-360). 
To avoid data leakage and ensure a fair evaluation, we ensure that the noise segments used in the training and test sets are non-overlapping. The resulting synthetic noisy data is then combined with the clean 
LibriSpeech data and shuffled to form the final training and validation sets.

Within our tri-stage training strategy, only the LibriSpeech clean dataset is used during the pre-training stage. In 
the subsequent noisy disentanglement training and fine-tuning stages, both clean and noisy speech are organized into 
clean-noisy and clean-clean training pairs to form the training dataset.

To validate the effectiveness of our approach under both simulated and real noisy conditions, we construct multiple 
noisy test sets. We select fixed SNRs from the set {-5, 0, 5, 10, 15} dB, using the test-clean subset of LibriSpeech as 
the clean base. Noise samples from the CHiME-4 dataset are then randomly selected to synthesize noisy test sets at 
these SNR levels. Specifically, 20 distinct simulated noisy test sets are created based on the four noise types (BUS, 
CAF, PED, STR) and the five SNR values, thereby evaluating the robustness of our method across various noisy 
conditions. Additionally, four real-noise test sets from the CHiME-4 Challenge, comprising recordings from the same 
four environments, are used to assess the performance under real noisy conditions.

To further evaluate generalization to unseen noise conditions, we construct an additional noisy test set based on the DEMAND 
dataset~\cite{thiemann2013demand}, which comprises 18 distinct types of real-world noise categorized into six groups: four 
inside or near-field environments (Domestic, Office, Public and Transportation) and two open-air environments (Street, Nature). Notably, none of these noise types are used during training, making the DEMAND dataset suitable for out-of-domain evaluation. We randomly select single-channel 16 kHz noise samples from DEMAND and mix them with utterances from the LibriSpeech test-clean 
subset at five fixed SNR levels (–5, 0, 5, 10, and 15 dB), forming the DEMAND simulated noisy test set for evaluating 
the model’s robustness under acoustically diverse and unseen noise conditions.

\subsection{Setups}
\label{subsec6}

The acoustic feature inputs used in our experiments are 80-dimensional log-mel filterbanks, 
which are normalized using global-level mean normalization. SpecAugment~\cite{park2019specaugment} is applied to 
augment the acoustic features during training. The vocabulary is composed of 5,000 byte-pair encoding (BPE) ~\cite{sennrich2015neural} 
units. It is important to note that the results presented in this study are only from the acoustic model, 
as no additional language model is incorporated during inference. 

For the encoder in Conformer-T model, we use a stack of 15 Conformer encoder layers. 
Each Conformer layer consists of a 1024-dimensional feed-forward module and a 256-dimensional attention 
module with 4 self-attention heads. For the Transducer Decoder, we use a single-layer, 256-dimensional 
LSTM-based RNN-T, thus resulting in a 30.53M model size. To facilitate model convergence, 
we apply a warmup mechanism with 15,000 steps, 
and a dropout rate of 0.1 is used for regularization to mitigate overfitting.  

During the model training, the scaling factor \(\mu\) and \(\gamma\) in Eq.(9) is set to 0.3 and 1. 
For the noisy disentanglement training and fine-tuning stages, the loss weights of $\alpha$ and \(\beta\) are set 
to 0.3 and 1, respectively. The Word Error Rate (WER) is used for evaluating our model's ASR performance.

\section{Results and Discussions}
\label{sec6}

\subsection{E2E ASR Baseline}
\label{subsec7}

\begin{table}[h!]%% placement specifier
\renewcommand\arraystretch{1.2}
    \centering
    \caption{WER\% of different Conformer-Transducer baselines on Librispeech dataset. 
    `Clean' and `Noisy' represent using train-clean-100 and train-noisy-100 dataset as model training, respectively. 
    B3 incorporates one additional Conformer encoder layer compared to B2. }
    \label{tab1}
    \vspace{0.5em}
    \setlength{\tabcolsep}{3pt} 
    \begin{tabular}{ccccccc} %% 表格列格式，| 用于竖线分隔
    \hline
    \multirow{2}{*}{\textbf{ID}} & \multirow{2}{*}{\textbf{Training Data}} & \multirow{2}{*}{\textbf{Params}} & \multicolumn{2}{c}{\textbf{Dev}} & \multicolumn{2}{c}{\textbf{Test}} \\ 
    \cline{4-5}   \cline{6-7} &  &  & \textbf{Clean} & \textbf{Other} & \textbf{Clean} & \textbf{Other} \\
    \hline
    B1 & Clean & 30.53 M & 9.3 & 23.0 & 9.5 & 23.7\\ %% A tabular row ends with \\
    B2 & Clean+Noisy & 30.53 M & 8.1 & 21.0 & 8.3 & 21.7\\
    B3 & Clean+Noisy & 32.12 M & 8.1 & 20.3 & 8.2 & 21.3\\\hline
\end{tabular}
\end{table}

Table \ref{tab1} summarizes the WERs of various Conformer-Transducer baselines trained on different 
datasets and with different parameter sizes. System B1 is trained only on the 100 hours of Librispeech 
clean data. System B2 and B3 trained on both 100 hours of clean data and 100 hours of simulated noisy data; 
B3, in particular, adds an additional Conformer encoder layer compared to B2, resulting in a similar 
model size to that of our proposed NoisyD-CT system.

As the table indicates, augmenting the training data with simulated noise (B2) substantially lowers the WER on both 
the clean and other test sets compared to B1, even though the model size remains unchanged. 
This improvement underscores  the benefits of increased training data diversity, even if a mismatch between 
training and testing conditions. Moreover, when comparing B2 and B3, the slight increase in model size 
(via one extra Conformer encoder layer) further boosts ASR performance on both the development and test sets.

\begin{table}[h!]
\renewcommand\arraystretch{1.2}
\centering
\caption{WER\% of different Conformer-Transducer baselines on simulated noisy test sets 
(taking the Librispeech test-clean as clean speech base) at different SNRs. 
Each SNR level covers BUS, CAF, PED, STR four noise types.}
\label{tab2}
\vspace{0.5em}
\setlength{\tabcolsep}{5pt}
\begin{tabular}{cccccccc} 
\hline
\multirow{2}{*}{\textbf{ID}} & \multirow{2}{*}{\textbf{Params}} & \multicolumn{5}{c}{\textbf{WER under SNR(dB)}} & \multirow{2}{*}{\textbf{Avg}}\\ 
\cline{3-7} 
 &  & \textbf{-5} & \textbf{0} & \textbf{5} & \textbf{10} & \textbf{15} \\
\hline
B2 & 30.53 M & 44.8	& 26.0 & 17.6 & 14.3 & 12.5 & 23.0\\ %% A tabular row ends with \\
B3 & 32.12 M & 43.5	& 24.7 & 16.0 & 12.5 & 10.9 & 21.5 \\
B3+Tri & 32.12 M & 44.8 & 26.0 & 16.3 & 12.2 & 10.6 & 22.0 \\
\hline
\end{tabular}
\end{table}

\begin{table*}[!ht]
\renewcommand\arraystretch{1.3}
\centering
\caption{WER\% of Conformer-T and the proposed NoisyD-CT with different training strategies. 
`ND' represents the noisy disentanglement module, `C-T' represents the Conformer-Transducer backbone. 
The pre-trained \textit{Feature Encoder} and \textit{Conformer Blocks} used to produce $h_{t}$ are 
frozen during both the noisy disentanglement training and fine-tuning 
stages. `w/o Encoder-N, Decoder-CN, $\mathcal{L}_{R}$' setting in S4 denotes a simplified NoisyD-CT variant 
where only Encoder-C and $\mathcal{L}_{CON}$ are used, without Encoder-N, Decoder-CN and $\mathcal{L}_{R}$. The training data of NoisyD-CT models is the same as the B3.}
\label{tab3}
\vspace{0.5em}
\setlength{\tabcolsep}{2.8pt}
\scalebox{0.85}{
\begin{tabular}{l|l|c|cc|cc|cc|cccc|ccccc}
\hline
\multirow{3}{*}{\textbf{ID}} & \multirow{3}{*}{\textbf{Model}} & \multicolumn{3}{c|}{\textbf{Noisy Disentanglement Training}} & \multicolumn{4}{c|}{\textbf{Fine-tuning}} & \multicolumn{4}{c|}{\textbf{LibriSpeech Test Sets}} & \multicolumn{5}{c}{\textbf{Noisy Test Sets}}\\ 
\cline{3-18} % 添加分隔线
 & & \textbf{Trainable} & \multicolumn{2}{c|}{\textbf{Loss}} & \multicolumn{2}{c|}{\textbf{Trainable}} & \multicolumn{2}{c|}{\textbf{Loss}} & \multicolumn{2}{c}{\textbf{Dev}} & \multicolumn{2}{c|}{\textbf{Test}} & \multicolumn{5}{c}{\textbf{WER under SNR(dB)}} \\
\cline{3-18}
 & & \textbf{ ND } & $\mathcal{L}_{Conformer-T}$ & $\mathcal{L}_{CON}+\mathcal{L}_{R}$ & \textbf{ C-T } & \textbf{ ND } & $\mathcal{L}_{Conformer-T}$ & $\mathcal{L}_{CON}+\mathcal{L}_{R}$ & \textbf{Clean} & \textbf{Other} & \textbf{Clean} & \textbf{Other} & \textbf{-5} & \textbf{0} & \textbf{5} & \textbf{10} & \textbf{15} \\
\hline
B3 & Conformer-T & - & - & - & - & - & - & - & 8.1	& 20.3 & 8.2 & 21.3 & 43.5 & 24.7 & 16.0 & 12.5 & 10.9\\

\hline
S1 & \multirow{4}{*}{NoisyD-CT} & \checkmark & \checkmark & \checkmark & \checkmark & \checkmark & \checkmark & \checkmark& 7.9 & 19.2 & 8.1 & 19.8 & 42.7 & 23.0 & 14.2 & 10.8 & 9.5 \\
S2  &  & \checkmark & - & \checkmark & \checkmark & - & \checkmark & - & 8.0 & 19.3 & 8.2 & 19.9 & 42.8 & 23.4 & 14.6 & 11.3 & 9.9 \\
S3  &   &  \checkmark & \checkmark & \checkmark & \checkmark & - & \checkmark & \checkmark & 8.0 &19.3 & 8.2 &19.7 & 41.8 & 22.2 & 13.5 & 10.2 & 9.0 \\
\cline{3-18}
S4 &  & \multicolumn{7}{c|}{w/o Encoder-N, Decoder-CN, $\mathcal{L}_{R}$} & 8.0	& 19.6 & 8.1 & 20.2 & 42.6 & 23.3 & 14.6 & 10.9 & 9.3\\
\hline
\end{tabular}}
%% Use \caption command for table caption and label.
\end{table*}

Table \ref{tab2} presents the WERs of the Conformer-Transducer baselines (B2, B3 and B3+Tri) on simulated noisy test sets 
across different SNR levels. The results show that incorporating an additional Conformer encoder layer in B3 
consistently improves performance over B2, reducing WER across all SNR conditions, particularly at higher SNR levels.
The reduction in WER is more pronounced in high-SNR conditions (e.g., 15 dB), 
suggesting that the increased model size is especially beneficial for learning cleaner acoustic patterns. 
However, the gap between B2 and B3 narrows at lower SNRs, indicating that the advantage of increased model 
size diminishes as noise levels rise, likely due to the model's limited ability in handling ASR under 
highly and challenging noisy environments. 

Moreover, beyond B3, we also build an additional baseline called B3+Tri to provide comprehensive system comparison with the proposed tri-stage trained NoisyD-CT. The only difference between B3 and B3+Tri lies the model training strategy. 
B3+Tri applies the similar tri-stage training schedule, as the proposed NoisyD-CT, while B3 is trained from scratch with random initialization. Specifically, the similar tri-stage training schedule of B3+Tri first pre-trained on clean speech under the same conditions as Stage 1 of NoisyD-CT, and then further trained on the combined clean-noisy speech dataset for a number of steps equivalent to Stages 2 and 3 as in NoisyD-CT. This baseline helps isolate the contribution of the training schedule itself from that of the proposed NoisyD-CT architecture and associated losses.
Compared to B3, the B3+Tri model shows marginal improvements at higher SNR levels such as 10 dB and 15 dB, but performs worse at lower SNR levels like –5 dB and 0 dB. Overall, the average WER increases 
from 21.5\% (B3) to 22.0\% (B3+Tri), indicating that the tri-stage training alone does not yield consistent or meaningful improvements. 
Since B3 achieves the best performance across various 
test conditions and has a model size comparable to our proposed NoisyD-CT model, we adopt it as 
the primary baseline for further comparisons.

\subsection{Ablation Results of Tri-stage Training and NoisyD components}
\label{subsec8}

Table \ref{tab3} compares the performance of our proposed NoisyD-CT model and the baseline system (B3) 
under various training strategies, evaluated on both the LibriSpeech test sets and simulated noisy speech at 
different SNR levels. The table is designed to illustrate how different approaches to 
noisy disentanglement training and fine-tuning affect overall speech recognition performance. 
Furthermore, to assess the effectiveness and necessity of noise suppression block (NoisyD) in the NoisyD-CT framework, we conducted an additional experiment based on the B3 baseline, with the Encoder-N, Decoder-CN and $\mathcal{L}_{R}$ of NoisyD are removed.
Specifically, B3 serves as the competitive baseline; S1 trains the noisy disentanglement module
using the full model loss while freezing the other modules, followed by a fine-tuning step on
the entire network (except the fixed \textit{pre-trained feature encoder and conformer blocks} used to produce
$h_{t}$). S2 similarly freezes the main modules while training the NoisyD module with only 
$\mathcal{L}_{CON}+\mathcal{L}_{R}$, then fine-tunes the Conformer-Transducer backbone using 
$\mathcal{L}_{Conformer-T}$. S3 implements the complete NoisyD-CT framework using our proposed 
tri-stage training strategy, as detailed in Section 4.3. S4 denotes a simplified NoisyD-CT variant 
where only Encoder-C and $\mathcal{L}_{CON}$ are used, while the Encoder-N, Decoder-CN and $\mathcal{L}_{R}$ are removed. 
Importantly, this variant is trained using the same tri-stage training strategy as S3, ensuring a fair comparison 
under identical training conditions.

The results demonstrate that NoisyD-CT outperforms B3 under all examined conditions, either on the 
LibriSpeech development and evaluation test sets, or on the simulated noisy test sets at different SNR levels, 
confirming the advantages of incorporating a noisy disentanglement mechanism and fine-tuning strategy.
Among the three variations (S1, S2, and S3), S3 yields the most substantial gains, particularly under 
higher-SNR  with relatively clean conditions. For example, the WER at a low SNR level decreases from 42.8\% to 41.8\%, 
while in cleaner conditions, the WER is reduced from 9.9\% to 9.0\%. 
Compared to B3, S4 yields moderate improvements on both clean and noisy test sets. For instance, the WER drops from 43.5\% to 42.6\% at –5 dB, and from 10.9\% to 9.3\% at 15 dB. However, this performance remains significantly inferior to the full NoisyD-CT model (S3). These results confirm that while Encoder-C, aided by the clean consistency loss, contributes to improved robustness, it is insufficient on its own.

Moreover, our full NoisyD-CT model (S3) consistently outperforms B3+Tri (shown in Table \ref{tab2}), across all SNR conditions. For example, at 0 dB SNR, WER is reduced from 26.0\% (B3+Tri) to 22.2\% (S3). These findings highlight not only the effectiveness of the proposed NoisyD-CT architecture itself but also the critical role played by the tri-stage training strategy in enhancing noise robustness and 
overall ASR performance.

\subsection{Noise Condition-wise Results}
\label{subsec9}

\begin{table*}[!ht]
\renewcommand\arraystretch{1.3}
\centering
\caption{WER\% of noise condition-wise results of competitive baseline B3 
and the proposed NoisyD-CT models.}
\label{tab4}
\vspace{0.5em}
\setlength{\tabcolsep}{3pt} 
\scalebox{0.9}{
\begin{tabular}{cc|ccccc|ccccc|ccccc|ccccc} 
\hline
\multirow{3}{*}{\textbf{ID}} & \multirow{3}{*}{\textbf{Model}} & \multicolumn{20}{c}{\textbf{WER under SNR(dB)}}\\ 
\cline{3-22} 
 & & \multicolumn{5}{c|}{\textbf{BUS}} & \multicolumn{5}{c|}{\textbf{CAF}} & \multicolumn{5}{c|}{\textbf{PED}} & \multicolumn{5}{c}{\textbf{STR}}\\
\cline{3-22}
 & & \textbf{-5} & \textbf{0} & \textbf{5} & \textbf{10} & \textbf{15} & \textbf{-5} & \textbf{0} & \textbf{5} & \textbf{10} & \textbf{15}& \textbf{-5} & \textbf{0} & \textbf{5} & \textbf{10} & \textbf{15}& \textbf{-5} & \textbf{0} & \textbf{5} & \textbf{10} & \textbf{15}\\
\hline
B3 & Conformer-T & 21.7 & 15.0 & 12.1 & 10.7 & 10.0 & 53.2 & 29.4 & 17.9 & 13.5 & 11.5 & 61.4 & 32.7 & 19.3 & 13.7 & 11.4 & 37.7 & 21.8 & 14.9 & 12.1 & 10.6\\ 
S3 & NoisyD-CT & 19.5 & 12.6 & 9.9 & 8.9 & 8.5 & 51.9 & 26.9 & 15.1 & 10.8 & 9.3 & 59.7 & 30.2 & 16.4 & 11.2 & 9.4 & 36.1 & 19.2 & 12.4 & 9.8 & 8.9\\
\hline
\end{tabular}}
\end{table*}

\begin{table*}[!ht]
\renewcommand\arraystretch{1.3}
\centering
\caption{WER\% comparison between the Conformer-Transducer and the proposed NoisyD-CT with tri-stage training on the LibriSpeech 360hrs trainset.}
\label{tab5}
\vspace{0.5em}
\scalebox{0.9}{
\begin{tabular}{cc|cccc|ccccc} 
\hline
\multirow{3}{*}{\textbf{ID}} & \multirow{3}{*}{\textbf{Model}} & \multicolumn{4}{c|}{\textbf{LibriSpeech}} & \multicolumn{5}{c}{\textbf{Noisy Test Sets}}\\
\cline{3-11}
&  & \multicolumn{2}{c}{\textbf{Dev}} & \multicolumn{2}{c|}{\textbf{Test}} & \multicolumn{5}{c}{\textbf{WER under SNR(dB)}}\\ 
\cline{3-11} % 添加分隔线
 & & \textbf{Clean} & \textbf{Other} & \textbf{Clean} & \textbf{Other} & \textbf{-5} & \textbf{0} & \textbf{5} & \textbf{10} & \textbf{15}\\
\hline
B3+ & Conformer-T & 7.4 & 16.6 & 7.9 & 16.7 & 33.2 & 18.0 & 13.0 & 11.4 & 10.9\\ 
S3+ & NoisyD-CT & 7.2 & 14.2 & 7.8 & 14.3 & 32.0 & 15.8 & 10.3 & 8.7 & 8.1 \\
\hline
\end{tabular}}
\end{table*}

Different from the results in Table \ref{tab2} and \ref{tab3}, Table \ref{tab4} 
presents a comparison of results between the competitive baseline B3 and our proposed NoisyD-CT S3 across 
five different SNR conditions in four different noise type environments. 
Overall, S3 outperforms B3 across all conditions, demonstrating better performance in noisy environments. 
Specifically, both B3 and S3 perform relatively better in the BUS noise environment, 
followed by the STR noise environment, while the CAF and PED noise environments yield the lowest performance. 
Comparing the results of B3 and S3, our proposed method consistently shows a similar improvement across all four noise 
environments at the same SNR level, with an enhancement of around 16.8\% to 20\% at SNR = 10dB. 
This could be due to the more uniform noise characteristics in the BUS environment, 
while the STR environment may include more diverse sounds like passing vehicles and pedestrians. 
Similarly, the CAF and PED environments contain a variety of low-frequency noise, making it more challenging to 
distinguish between speech and noise, thereby reducing the system's performance. 
Despite these challenges, our proposed NoisyD-CT still improves the noise robustness of the speech recognition 
system in various scenarios.

By comparing the performance of S3 and B3 under different SNR conditions within the same noise environment, 
it is evident that the higher SNR, the more WER decreases. This is understandable because a higher SNR means 
that the environment or application scenarios are relatively clean or quiet, reducing interference and allowing 
the model to better distinguish clear speech, thus improving ASR performance. 
Notably, under high SNR (15 dB), the interference from noise is greatly reduced, and the model performance 
shows a marked improvement. All these findings show that our proposed NoisyD-CT framework demonstrates 
excellent performance across all noise environments, 
indicating its superior noise robustness over the standard Conformer-Transducer model.

\subsection{Results with 360 hrs LibriSpeech Source}
\label{subsec10}

Table \ref{tab5} demonstrates the performance of both the competitive Conformer-T baseline and our proposed NoisyD-CT with 
tri-stage training, each trained on a larger dataset comprising 360-hours of LibriSpeech clean speech (train-clean-360) and 
360-hours of simulated noisy speech (train-noisy-360). These configurations are denoted as B3+ and S3+ in the table. 
Notably, NoisyD-CT (S3+) outperforms the baseline across the LibriSpeech test sets and various SNR conditions. On the clean 
test sets, the WER decreases from 7.9\% to 7.8\% on test-clean, and from 16.7\% to 14.3\% on test-other, reflecting 
consistent improvements also observed on dev-clean and dev-other. The larger WER reductions for dev-other and test-other 
suggest that NoisyD-CT more effectively handles noisy or challenging acoustic conditions.

In addition, on the simulated noisy test sets, S3+ achieves at least a 3.6\% relative WER reduction at 
SNR = -5, with improvements up to 25.7\% at SNR = 15. These gains highlight NoisyD-CT’s ability to retain or even enhance 
performance under clean conditions while substantially improving ASR performance in noisy environments. 
Moreover, when compared to the 100-hour training results in Tables \ref{tab1} and \ref{tab2}, the larger 360-hour dataset 
yields nearly double the relative reduction in WER, indicating stronger generalization and robustness to noise.

\begin{figure*}[!ht]
    \centering
    \includegraphics[width=1\linewidth]{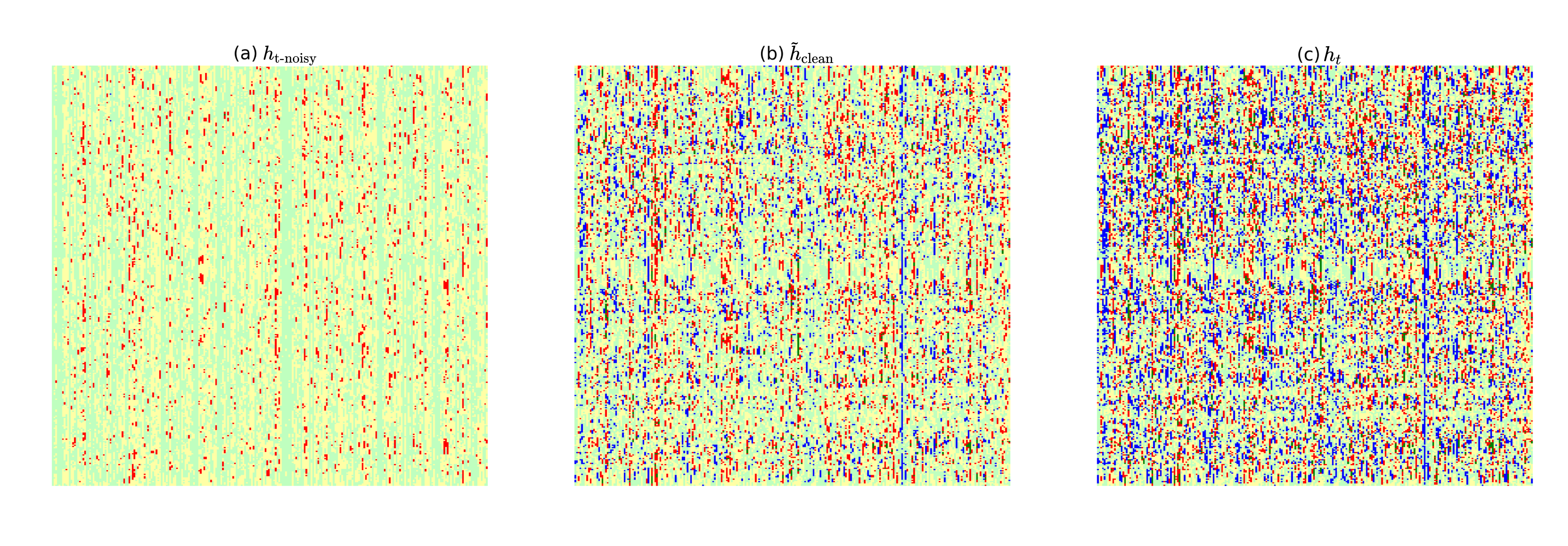}
    \caption{Visualization of the (a) noisy representation $h_{t-noisy}$ , (b) 
    the disentangled clean representation \(\tilde{h}_{clean}\), and (c) the ground-truth clean representation $h_t$ .}
    \label{fig4}
\end{figure*}

\begin{table}[!htbp]
\renewcommand\arraystretch{1.3}
\centering
\caption{WER\% on the CHiME-4 Challenge real-world noisy test sets with models of B3+ and S3+.} 
\label{tab6}
\vspace{0.5em}
\setlength{\tabcolsep}{8pt} % 减小列间距
\begin{tabular}{cccccc} %% 表格列格式，| 用于竖线分隔
\hline
\multirow{2}{*}{\textbf{ID}} & \multirow{2}{*}{\textbf{Model}} & \multicolumn{4}{c}{\textbf{Real-noisy Test Sets}}\\ 
\cline{3-6} % 添加分隔线
 & & \textbf{BUS} & \textbf{CAF} & \textbf{PED} & \textbf{STR}\\
\hline
B3+ & Conformer-T & 39.6 & 34.3 & 32.3 & 29.2 \\ 
S3+ & NoisyD-CT & 38.3 & 31.3 & 30.4 & 26.1 \\
\hline
\end{tabular}
\end{table}

Table \ref{tab6} presents the performance of our proposed NoisyD-CT with tri-stage training mechanisim on the CHiME-4 real-world noisy test sets. In the BUS noise environment, the WER decreases slightly from 39.6\% to 38.3\%. In contrast, in the STR noise environment, our NoisyD-CT achieves a remarkable 10.6\% relative WER reduction than the Conformer-Transducer baseline. This demonstrates that our method remains effective in real-world noisy environments. However, the extent of performance improvement varies significantly depending on the type of noise environment, which may be attributed to the inherent differences in speech characteristics across different noise types.

\begin{table}[!htbp]
\renewcommand\arraystretch{1.3}
\centering
\caption{WER\% on the DEMAND simulated noisy test sets with models of B3+ and S3+.} 
\label{tab7}
\vspace{0.5em}
\setlength{\tabcolsep}{6pt} % 减小列间距
\begin{tabular}{ccccccc} %% 表格列格式，| 用于竖线分隔
\hline
\multirow{2}{*}{\textbf{ID}} & \multirow{2}{*}{\textbf{Model}} & \multicolumn{5}{c}{\textbf{WER under SNR(dB)}}\\ 
\cline{3-7} % 添加分隔线
 & & \textbf{-5} & \textbf{0} & \textbf{5} & \textbf{10} & \textbf{15}\\
\hline
B3+ & Conformer-T & 19.4 & 13.6 & 11.6 & 10.9 & 10.6 \\ 
S3+ & NoisyD-CT & 17.0 & 11.0 & 8.8 & 8.2 & 7.9 \\
\hline
\end{tabular}
\end{table}

Table \ref{tab7} presents WERs of the baseline Conformer-Transducer model (B3+) and our proposed NoisyD-CT model (S3+) 
on the DEMAND simulated noisy test sets across five fixed SNR levels ranging from the set -5, 0, 5, 10, 15 dB. Notably, the DEMAND 
dataset contains entirely unseen noise types during training, making this a strong benchmark for evaluating generalization. 
Across all SNR conditions, NoisyD-CT (S3+) consistently outperforms the baseline (B3+), demonstrating substantial 
improvements in noise robustness. At –5 dB, where the noise is most severe, S3+ reduces the WER from 19.4\% to 17.0\%, 
a relative improvement of 12.4\%. At higher SNRs, the observed improvements are more prominent and significant. For instance, at 15 dB, S3+ achieves a WER of 7.9\%, compared to 10.6\% for B3+, marking a relative improvement of 25.5\%. These results 
clearly demonstrate that our proposed NoisyD-CT framework not only performs well under matched noisy conditions, but also generalizes effectively to completely unseen and acoustically diverse noise types.

\subsection{Visualization}
\label{subsec11}

Figure \ref{fig4} illustrates the ability of our proposed NoisyD-CT framework to disentangle noisy 
speech into a clean representation. The example shown corresponds to a simulated noisy utterance with ID 
 \texttt{26-496-0021.wav}, containing STR noise at a SNR of 7. The three subfigures are: (a) the high-
 dimensional noisy representation $h_{t-noisy}$, output by the backbone Conformer blocks when processing 
 the noisy input;  (b) the disentangled clean representation  \(\tilde{h}_{clean}\), generated by the 
Encoder-C of the NoisyD-CT model; and (c) the ground-truth clean representation $h_t$, obtained from the 
pre-trained Feature Encoder and Conformer blocks when the corresponding clean speech is provided as input.

From Figure \ref{fig4}, the visual comparison reveals that the disentangled clean representation
\(\tilde{h}_{clean}\) in (b) closely aligns with the clean representation of ground-truth  $h_t$ 
in (c), indicating that the noise has been effectively suppressed while preserving the acoustic 
characteristics of clean speech. This strong similarity highlights the effectiveness of our proposed  
NoisyD-CT model’s noisy disentanglement mechanism and provides additional evidence of its ability to 
improve ASR performance under noisy conditions.

\section{Conclusion}
\label{sec7}

In this paper, we introduced NoisyD-CT, an end-to-end framework designed to enhance the noise 
robustness of the state-of-the-art Conformer-Transducer speech recognition system. 
The core of this framework is a lightweight noisy disentanglement module, placed between the final Conformer blocks and 
the Transducer Decoder, which increases the model size by only a small margin (approximately 1.71M parameters). 
To train this module effectively, we proposed a tri-stage training approach that enables the extraction of cleaner 
representations from noisy inputs. Experimental results on both simulated noisy test sets with varying SNR levels 
and real-world CHiME-4 data demonstrate significant ASR performance gains. Notably, the proposed NoisyD-CT framework 
preserves, and in some cases even improves ASR performance on clean speech compared to the baseline Conformer-Transducer 
model trained only on clean training data. This highlights NoisyD-CT’s robust noise suppression ability and strong 
generalization without sacrificing performance on clean inputs. Our future work will focus on 
generalizing our proposed methods on larger-scale noisy ASR datasets and tasks.

%% Use figure environment to create figures
%% Refer following link for more details.
%% https://en.wikibooks.org/wiki/LaTeX/Floats,_Figures_and_Captions
% \begin{figure}[t]%% placement specifier
% %% Use \includegraphics command to insert graphic files. Place graphics files in 
% %% working directory.
% \centering%% For centre alignment of image.
% \includegraphics{example-image-a}
% %% Use \caption command for figure caption and label.
% \caption{Figure Caption}\label{fig1}
% %% https://en.wikibooks.org/wiki/LaTeX/Importing_Graphics#Importing_external_graphics
% \end{figure}

% \appendix
% \section{Example Appendix Section}

%%       \cite{<label>} ==> [1]

%%

\bibliographystyle{elsarticle-num} 
\bibliography{main}



\end{document}